\documentclass[journal,letterpaper]{IEEEtran}

\usepackage{amssymb}
\usepackage{amsthm}
\usepackage{amsmath}
\usepackage{quantum}
\usepackage{hyperref}
\usepackage{tikz}

\usetikzlibrary{}
\usepgfmodule{oo}

\allowdisplaybreaks[1]

\begin{document}

\title{Polar Codes in Network Quantum Information Theory}

\author{Christoph Hirche, Ciara Morgan,  Mark M. Wilde%
\thanks{C. Hirche is with F\'{\i}sica Te\`{o}rica: Informaci\'{o} i Fen\`{o}mens
 Qu\`{a}ntics, Departament de F\'{i}sica, Universitat Aut\`{o}noma de Barcelona, ES-08193
 Bellaterra (Barcelona), Spain. C. Morgan is with the School of Mathematics and Statistics, University College Dublin, Belfield, Dublin 4. Ireland. C. Hirche and C. Morgan were with the Institut f\"{u}r Theoretische Physik, Leibniz Universit\"{a}t Hannover, Appelstra\ss e 2, D-30167 Hannover, Germany (e-mail: \{christoph.hirche, ciara.morgan\}@itp.uni-hannover.de). M. M. Wilde  is with the Hearne Institute for Theoretical Physics, Department of Physics and Astronomy,  Center for Computation and Technology,  Louisiana State University, Baton Rouge, Louisiana 70803, USA. (e-mail: mwilde@lsu.edu)}
\thanks{C. Hirche acknowledges support by the Spanish MINECO, project FIS2013-40627-P, as well as by the Generalitat de Catalunya, CIRIT project no. 2014 SGR 966. C. Hirche and C. Morgan acknowledge support from the EU grants SIQS and QFTCMPS and from the cluster of excellence EXC 201 Quantum Engineering and Space-Time Research. M. M. Wilde acknowledges startup funds from the Department of Physics and
Astronomy at LSU, support from the NSF through Award No.~CCF-1350397, and
support from the DARPA Quiness Program through US Army Research Office award
W31P4Q-12-1-0019.}}

\date{\today}
\maketitle

\begin{abstract}
Polar coding is a method for communication over noisy classical channels
which is provably capacity-achieving and has an efficient encoding and
decoding. Recently, this method has been generalized to the realm of quantum
information processing, for tasks such as classical communication, private
classical communication, and quantum communication. In the present work, we
apply the polar coding method to network classical-quantum information theory, by
making use of recent advances for related classical tasks. In particular, we
consider problems such as the compound multiple access channel and the
quantum interference channel. The main result of our work is that it is
possible to achieve the best known inner bounds on the achievable rate
regions for these tasks, without requiring a so-called quantum simultaneous
decoder. Thus, this paper paves the way for developing network classical-quantum
information theory further without requiring a quantum simultaneous decoder.
\end{abstract}


\section{Introduction}

\IEEEPARstart{O}{ne} of the key tasks distinguishing the theory of quantum physics from its
classical counterpart is the simultaneous measurement of non-commuting
observables. Indeed, the uncertainly principle captures one of the most
profound characteristics of quantum mechanics, that is, the impossibility of
simultaneously measuring non-commuting operators to arbitrary accuracy. The
principle itself is considered a cornerstone of modern physics.

In quantum communication theory, the problem of simultaneous measurement
arises in multi-user communication models when one needs to simultaneously
learn about two or more non-commuting output states of a quantum channel in
order to achieve optimal rates of communication. In particular, the problem
of simultaneous decoding in the quantum setting manifests itself in the
difficulty in constructing a measurement operator achieving this task.
Although this problem is well understood classically, the existence of a
general quantum simultaneous decoder has remained a conjecture.

In contrast to simultaneous measurement, building decoders based on
measuring outputs of two or more users of a channel successively, i.e.,
successive decoding, has been successfully realized in the quantum setting 
\cite{W01}. Moreover, a coding strategy based on successive decoding has
been shown to achieve the optimal communication rate region for the
classical-quantum multiple access channel (cq-MAC) \cite{W01}, with the help
of the gentle measurement lemma \cite{W99} bounding the measurement
disturbance of quantum states.
\IEEEpubidadjcol

Despite this success with successive decoding and the multiple access
channel, achieving the rate region for other multi-user quantum channels has
remained elusive. An important example of such a channel is provided by the
quantum interference channel \cite{FHSSW12}. In this model two or more
senders wish to communicate information simultaneously, and solely with
their intended corresponding receiver by means of a noisy channel modeled by
cross-talk or interference. In the classical setting, the capacity of the
interference channel is known exactly only in the case of very strong
interference~\cite{Carleial75} or strong interference~\cite{HK81}.

The best known achievable rate region for the two-user classical
interference channel is given by the Han-Kobayashi region \cite{HK81}. The
coding strategy which achieves this region relies on the simultaneous
decoding of two three-sender multiple access channels. A quantum version of
the Han-Kobayashi rate region for the classical-quantum interference channel
has also been conjectured to be achievable, based on the conjectured
existence of a three-sender simultaneous decoder \cite{FHSSW12, SFWSH11},
and it was in fact proven to be achievable in \cite{S11} using a specialized
three-sender quantum simultaneous decoder. This result raised the question
of whether the quantum Han-Kobayashi region can be achieved using a
successive decoder \cite{FS12}.

\smallskip

In this article, we address this question by exploiting Arikan's polar
coding technique for classical channels \cite{A09}. Indeed, polar codes have
attracted a great deal of attention as the first \emph{constructive}
capacity-achieving codes with an efficient encoder and decoder. Recently the
polar coding technique has been applied to a variety of multi-user classical
channels including multiple access channels \cite{STY10, STY13, O13, MELK13}%
, broadcast channels \cite{GAG13, MHSU14}, interference networks \cite{WS14}
and for the task of source coding \cite{A12} and universal coding for
compound channels \cite{HU13, MELK132, WS14}. We make use of some of these
advances in our work.  Polar coding has also been generalized for the task
of sending classical \cite{WG13,WG12} and quantum information \cite{WR12,
WG13QDeg} over single-user quantum channels. In the quantum setting,
efficiency has only been shown in general for the encoder and left open for
the decoder \cite{WG13,WG12,WR12} (see \cite{WLH13} for recent progress on
the efficient decoder question). However an efficient encoding and decoding
scheme has been shown for certain quantum channels in the case of quantum
communication \cite{RDR12}.

In this work we show that polar coding can also be applied to the cq-MAC to
achieve every point in the known achievable rate region \cite{W01} and also that an
approach for universal polar codes from \cite{HU13} can be used to obtain
achievable rates for compound cq-MACs. We also apply the results
obtained for compound cq-MACs in a way similar to \cite{WS14} in order to achieve the
Han-Kobayashi rate region for the two-user classical-quantum interference
channel using a successive cancellation decoder.

The paper is organized as follows. In Section \ref{prel} we introduce the
necessary mathematical preliminaries and the classical-quantum multiple
access and interference channels, along with the technique for polar coding
for classical-quantum channels in the case of a single sender and receiver.
In Section \ref{2MAC} we discuss the two-user classical-quantum MAC and
generalize in Section \ref{3MAC} to multiple users and in Section \ref{cMAC}
to compound MACs. This is applied to interference channels in Section \ref%
{int}, and we conclude in Section \ref{concl} with a summary and some open
questions.

\section{Preliminaries}

\label{prel}

\subsection{Notation and definitions}

A discrete classical-quantum channel $W$ takes realizations $x \in \mathcal{X%
}$ of a random variable $X$ to a quantum state, denoted $\rho_x^B$, on a
finite-dimensional Hilbert space $\mathcal{H}^B$, 
\begin{equation}
W : x \rightarrow \rho_{x}^{B},
\end{equation}
where each quantum state $\rho_x$ is described by a positive semi-definite
operator with unit trace. We will take the input alphabet $\mathcal{X} =
\{0,1\}$ unless otherwise stated, and the tensor product $W^{\otimes N}$ of $%
N$ channels is denoted by $W^N$.

To characterize the behavior of the channels, we will make use of the
symmetric Holevo capacity, defined as follows: 
\begin{equation}
I(W) \equiv I(X;B)_\rho,
\end{equation}
where the quantum mutual information with respect to a classical-quantum
state $\rho^{XB}$ is given by  
\begin{equation}
I(X;B) \equiv H(X)_\rho + H(B)_\rho - H(XB)_\rho,
\end{equation}
with $\rho^{XB} = \frac{1}{2} | 0\rangle\!\langle 0 | \otimes \rho_0^B + 
\frac{1}{2} | 1\rangle\!\langle 1 | \otimes \rho_1^B$. In the above, the von
Neumann entropy $H(\rho)$ is defined as 
\begin{equation}
H(\rho) \equiv -\tr\{\rho \log_2 \rho\}.
\end{equation}
We will also make use of the quantum conditional mutual information defined
for a tripartite state $\rho^{XYB}$ as 
\begin{multline}
I(X;B|Y)_\rho \equiv H(XY)_\rho + H(YB)_\rho - H(Y)_\rho \\ - H(XYB)_\rho.
\end{multline}
We characterize the reliability of a channel $W$ as the fidelity between the
output states 
\begin{equation}
F(W) \equiv F(\rho_0, \rho_1),
\end{equation}
with 
\begin{equation}
F(\rho_0, \rho_1) \equiv \normTr{\sqrt{\rho_0}\sqrt{\rho_1}}^2
\end{equation}
and 
\begin{equation}
\normTr{A} \equiv \tr{\sqrt{A^\dagger A}}.
\end{equation}
The Holevo capacity and the fidelity can be seen as quantum generalizations
of the mutual information and the Bhattacharya parameter from the classical
setting, respectively (see, e.g., \cite{A09}).

\subsection{Classical-quantum multi-user channels}

\label{cqMulti}

In the following sections, we will focus on two particular kinds of
multi-user channels: classical-quantum multiple access channels (cq-MACs)
and classical-quantum interference channels.

We begin with the classical-quantum interference channel, and for simplicity
we focus on the case of
two senders and two receivers.
The interference channel can
be modeled mathematically as the following triple: 
\begin{equation}
\left( \mathcal{X}_{1}\times \mathcal{X}_{2},W,\mathcal{H}^{B_{1}}\otimes 
\mathcal{H}^{B_{2}}\right) ,
\end{equation}%
with 
\begin{equation}
W:x_{1},x_{2}\rightarrow \rho _{x_{1},x_{2}}^{B_{1}B_{2}}.
\end{equation}
The information
processing task for the classical-quantum interference channel \cite%
{FHSSW12,SFWSH11} is as follows. The $k$th sender
would like to communicate a message to the $k$th
receiver, where $k \in \{1,2\}$. Sender $k$ chooses a message $m_k$ from a message set $\mathcal{M}%
_k = \{1, \cdots, 2^{N R_k} \}$, and encodes her message as a codeword $%
x_k^N(m_k) \in \mathcal{X}_k^N$.
The encoding for each sender is given by $%
\{x_{1}^{N}(m_{1})\}_{m_{1}\in \mathcal{M}_{1}}$ and $\{x_{2}^{N}(m_{2})%
\}_{m_{2}\in \mathcal{M}_{2}}$, respectively, with the corresponding
receivers' decoding POVMs denoted by $\{\Lambda _{m_{1}}\}$ and $\{\Gamma _{m_{2}}\}$.
The code is said to be an $(N,R_{1},R_{2},\epsilon )$-code, if the average
probability of error is bounded as follows 
\begin{equation}
\bar{p}_{e}=\frac{1}{|\mathcal{M}_{1}||\mathcal{M}_{2}|}%
\sum_{m_{1},m_{2}}p_{e}(m_{1},m_{2})\leq \epsilon ,
\end{equation}%
where the probability of error $p_{e}(m_1, m_2)$ for a pair of messages $(m_{1},m_{2})$
is given by 
\begin{equation}
p_{e}(m_{1},m_{2})=\mathrm{Tr}\left\{\left( I-\Lambda _{m_{1}}\otimes \Gamma
_{m_{2}}\right) \rho _{x_{1}^{N}(m_{1}),x_{2}^{N}(m_{2})}^{B^N_{1}B^N_{2}}\right\},
\end{equation}
with $\rho _{x_{1}^{N}(m_{1}),x_{2}^{N}(m_{2})}^{B^N_{1}B^N_{2}}$ the state resulting
when senders 1 and 2 transmit the codewords $x_{1}^{N}(m_{1})$
and $x_{2}^{N}(m_{2})$ through $N$ instances of the channel, respectively.

A rate pair $(R_1, R_2)$ is said to be \emph{achievable} for the two-user
classical-quantum interference channel described above if there exists an $%
(N, R_1,R_2, \epsilon)$-code $\forall \epsilon >0$ and sufficiently large~$N$.

The two-user classical-quantum interference channel induces two c-q MACs
which can be modeled as 
\begin{equation}
\left( \mathcal{X}_{1}\times \mathcal{X}_{2},\rho _{x_{1},x_{2}}^{B_{1}}=%
\mathrm{Tr}_{B_{2}}\{\rho _{x_{1},x_{2}}^{B_{1}B_{2}} \},\mathcal{H}%
^{B_{1}}\right) ,
\end{equation}%
and 
\begin{equation}
\left( \mathcal{X}_{1}\times \mathcal{X}_{2},\rho _{x_{1},x_{2}}^{B_{2}}=%
\mathrm{Tr}_{B_{1}} \{\rho _{x_{1},x_{2}}^{B_{1}B_{2}}\},\mathcal{H}%
^{B_{2}}\right) .
\end{equation}

The \emph{rate region} for a channel is given by the closure of all
achievable rates for that channel. We will be particularly interested in the
Han-Kobayashi rate region for the two-user interference channel. This region
was achieved in the classical setting by exploiting a coding strategy for
the interference channel which induces two three-user MACs,
together with a simultaneous decoder~\cite{HK81}.

The two-user c-q MAC is defined by the following triple, corresponding to the
input alphabets, channel output state and output system:
\begin{equation}
\left( \mathcal{X}_{1}\times \mathcal{X}_{2},\rho _{x_{1},x_{2}}^{B},\mathcal{H}^{B}\right) .
\end{equation}
The coding task is for two senders to communicate individual messages to a
single receiver. The detailed description of the information processing task is
somewhat similar to the above, so we omit it for brevity's sake.

Later, we will be particularly interested in \emph{compound} cq-MACs. Indeed, compound channels form a class
of channels with so-called ``channel uncertainty.'' In this model, a channel
is chosen from a set of possible channels, and used to transmit the
information, thus generalizing the traditional setting in which both sender
and receiver have full knowledge of the channel before choosing their code.
The classical and quantum capacities of compound quantum channels have been
studied in \cite{BBN08, BB09}, respectively. 

A compound cq-MAC is defined by a set $\mathcal{W} = \{ W_i\}$ of cq-MAC channels where each $W_i$ can be written as
\begin{equation}
W_i : x_1 , x_2 \rightarrow \rho_{x_1,x_2,i}^{B},
\end{equation}
and characterized by its output state $\rho_{x_1,x_2,i}^{B}$, taken with respect to the input pairs $(x_1,x_2) \in \mathcal{X}_1 \times \mathcal{X}_2$.

Note that we will look at the case in which the receiver knows the particular channel $W_i$ which has been chosen. However, the sender does not have this knowledge. This assumption can be easily justified in the case of taking many uses of the channel, since channel tomography can be performed in order to give the receiver knowledge of the channel which has been chosen. Moreover, this requires a small number of channel uses when compared to the overall number of channel uses, thereby not affecting the communication rate. 

\subsection{Polar codes for classical-quantum channels}\label{cq}

In \cite{WG13} a polar coding scheme for single-user classical-quantum
channels was introduced. We review the scheme briefly below before applying
the technique to the c-q MAC and interference channel in the following
sections.

Polar codes exploit the effect of channel polarization, which is achieved in
two steps, namely, by so-called channel combining and channel splitting.
In \textit{channel combining} the input sequence $u^N$ is transformed by a
linear transformation given by $x^N=u^NG_N$ where 
\begin{equation}
G_N = B_NF^{\otimes n}
\end{equation}
with $N = 2^n$,
\begin{equation}
F \equiv \left[ 
\begin{matrix}
1 & 0 \\ 
1 & 1%
\end{matrix}%
\right],
\end{equation}
and $B_N$ is a permutation matrix known as a ``bit reversal'' operation \cite%
{A09}. This induces a channel $W_N$ from $N$ single copies of the channel $W$%
.
For the \textit{channel splitting} step the combined channel $W_N$ from the
previous step is used to define new channels $W^{(i)}_N$ as follows:
\begin{equation}
W^{(i)}_N : u_i \rightarrow \rho_{(i),u_i}^{U_1^{i-1}B^N},
\end{equation}
where 
\begin{equation}
\rho^{U^{i-1}_1 B^N}_{(i),u_i} = \sum_{u_1^{i-1}} \frac{1}{2^{i-1}} %
\ketbra{}{u_1^{i-1}}{u_1^{i-1}} \otimes \sum_{u_{i+1}^N} \frac{1}{2^{N-i}}
\rho^{B^N}_{u^N}
\end{equation}
and $u_1^N$ denotes a row vector $(u_1, \dots, u_N)$ and correspondingly $%
u_i^j$ denotes, for $1\leq i, j\leq N$, a subvector $(u_i, \dots, u_j)$.
Note that if $j<i$ then $u_i^j$ is empty. Similarly for a vector $u_1^N$ and
a set $A \subset \{ 1,\dots, N\}$ we write $u_A$ to denote the subvector $%
(u_i : i\in A)$.
The above can be seen as a ``genie-aided'' successive cancellation decoder,
where the $i$-th measurement estimates the bit $u_i$, with the
assumptions that the entire output is available to the decoder, the previous
bits $u_1^{i-1}$ are correctly decoded and the distribution over the bits $%
u^N_{i+1}$ is uniform.

The channel polarization effect ensures that the fraction of channels $%
W_N^{(i)}$ which have the property $I(W_N^{(i)}) \in (1-\delta, 1]$ goes to
the symmetric Holevo information $I(W)$ and the fraction with $I(W_N^{(i)})
\in [0, \delta)$ goes to $1-I(W)$ for any $\delta \in (0,1)$, as $N$ goes to
infinity through powers of two \cite{A09,WG13} (see \cite{WG13} for
a precise statement). Hence we choose a polar code as a ``$G_N$%
-coset code'' \cite{A09}; that is, we choose a subset $A \subset
\{1,\dots,N\}$ and re-write the input transformation 
\begin{equation}
x^N=u^NG_N
\end{equation}
as 
\begin{equation}
x^N=u_AG_N(A) \oplus u_{A^c} G_N(A^c),
\end{equation}
where $G_N(A)$ denotes the submatrix of $G_N$ constructed from the rows of $%
G_N$ with indices in $A$. Now we can fix a code $(N,K,A,u_{A^c})$ where $N$
is the length of the code, $K=|A|$ is the number of information bits, $A$
fixes the indices for the information bits and $u_{A^c}$ is the vector of
so-called frozen bits.

A polar code has the above properties and is such that it obeys the
polar coding rule, which is that the set of information indices $A$ is
chosen such that the following inequality holds between fidelities 
\begin{equation}
F(W^{(i)}_N) \leq F(W^{(j)}_N)
\end{equation}
for all $i\in A$ and $j\in A^c$.

Lastly, a bound on the block error probability $P_e(N,R)$ for blocklength $N$ and rate $R$ was derived for a fixed $%
R<I(W)$ and $\beta< \frac{1}{2}$, with the result \cite{WG13} 
\begin{equation}
P_e(N,R) = o(2^{-\frac{1}{2}N^\beta}).
\end{equation}
The measurement achieving this error bound was called a quantum successive
cancellation decoder \cite{WG12}, and the error analysis exploited Sen's non-commutative union bound \cite{S11}.

The technique described in this section can also be extended to non-binary alphabets using the approach in \cite{STA09}, where an additional permutation is introduced to the coding process, which ensures polarization. Since these permutations only act on the inputs of the channel, which are also in our setting classical, this approach can be directly translated to classical-quantum channels. For further details we refer to \cite{STA09}. 

\section{Polar codes for the two-user binary-input cq-MAC}

\label{2MAC}

The achievable rate region for the classical-quantum MAC is
described by the following bounds~\cite{W01}: 
\begin{align}
R_{x} &\leq I(X;B|Y)_\rho , \\
R_{y} &\leq I(Y;B|X)_\rho , \\
R_{x}+R_{y} &\leq I(XY;B)_\rho ,
\end{align}%
with respect to a ccq-state 
\begin{equation}
\rho ^{XYB}=\sum_{x,y}p_{X}(x)p_{Y}(y)\ketbra{X}{x}{x}\otimes %
\ketbra{Y}{y}{y}\otimes \rho _{x,y}^{B}.
\end{equation}

The case in which the last inequality above is saturated is of particular
interest to us and the resulting line, which interpolates between the points 
$(I(X;B)_\rho, I(Y;B|X)_\rho )$ and $(  I(X;B|Y)_\rho,I(Y;B)_\rho )$, is called the \textit{%
dominant face} of the rate region. It is clear that if every point on the
dominant face can be achieved then we can also achieve every other point
within the rate region by resource wasting.

Recently, Arikan introduced the technique of ``monotone chain rules''
for handling the Slepian-Wolf problem
\cite{A12} with the polar coding technique, and
\"Onay applied this approach to the binary-input MAC \cite{O13}.
The advantage of this approach is that with each monotone chain rule,
we can achieve a rate pair lying on the dominant face of the rate region.
Furthermore, the achievable points form a dense subset of all points
on the dominant face, so that we can approximate every point on the dominant face to
arbitrarily good accuracy. Note that it will be crucial when coding for the interference channel that the compound MAC described later can achieve specific points on the dominant face of the rate region, which is ensured by the approach described in this section. 
Here we apply the technique to the classical-quantum MAC with two senders, each with binary alphabet.

The main difficulty in generalizing Arikan's technique to the classical-quantum setting lies in adjusting the decoder for the quantum outputs of the channel, such that they properly handle messages from two different senders while keeping the exponentially decreasing error. In addition we generalize the proofs for the continuity of the achievable rates and the path scaling to the classical-quantum setting. 

We now recall the idea of a monotone chain rule, but with our discussion here being for
the classical-quantum MAC. 
Let $X^N$ and $Y^N$ each denote a sequence of $N$ uniformly random bits. Let
$U^N$ be the result of sender~1 processing the sequence $X^N$ with the polar encoder,
and let $V^N$ be the result of sender~2 processing the sequence $Y^N$ with the polar encoder.
Let $(S_1, \dots, S_{2N})$ be a permutation
of the input sequence $U^NV^N$ such that the relative order of the elements
constituting $U^N$ is preserved. A chain-rule expansion for mutual
information is said to be monotone with respect to $U^N$ if it is of the
following form:
\begin{align}  
N \cdot I(XY;B) & = I(U^N V^N;B^N) \\
& = \sum_{i=1}^{2N} I(S_i;B^N | S^{i-1}), \label{NI}
\end{align}
with the first equality following from the reversibility of the encoders and the
second from the chain rule for mutual information.
Based on the above permutation, we let $b^{2N}$ denote a binary sequence which
we can think of as a ``path,'' where $b_k$ is equal to zero
if the $k$th channel use is transmitting an information bit from the input
sequence $U^N$ of the first sender and equal to one if the $k$th channel use
is transmitting an information bit
from the input sequence $V^N$ of the second sender. This gives rise to the
following rates:
\begin{align}
R_x &= \frac{1}{N} \sum_{k:b_k=0} I(S_k;B^N|S^{k-1}) \nonumber\\
&\leq \frac{1}{N} I(U^N ;B^N|V^N) = I(X;B|Y) ,\\
R_y &= \frac{1}{N} \sum_{k:b_k=1} I(S_k;B^N|S^{k-1}) \nonumber\\
&\leq \frac{1}{N} I(V^N ;B^N|U^N) = I(Y;B|X) ,\\
R_x + R_y &= I(XY;B),
\end{align}
where the inequalitites hold because of the structure of the monotone chain
rules in (\ref{NI}), the statistical independence of $U^N$ and $V^N$,
and the one-to-one correspondence between $U^N, V^N$ and $X^N, Y^N$,
respectively. 

To illustrate the chain rule approach, consider for example the case where we limit the form of the chain rule to splitting the first sender's messages into two parts corresponding to the chain $b^{2N} =  0^j 1^N 0^{N-j}$. Here we would get the explicit rates
\begin{align}
R_x &= \frac{1}{N} \sum_{i=1}^j I(U_i;B^N|U^{i-1}) \nonumber \\
& \ \ \ \ +  \frac{1}{N} \sum_{i=j+1}^N I(U_i;B^N|U^{i-1}V^N), \\
R_y &= \frac{1}{N} \sum_{i=1}^N I(V_i;B^N|U^jV^{i-1}),
\end{align}
which achieve points on the dominant face corresponding to the particular choice of the parameter $j$. 

Using the standard polar coding channel-combining technique
outlined in the previous section, we get a combined channel $W_N$ from $W^N$ by transforming
both input sequences as
\begin{equation}
x^N = u^N G_N, \qquad y^N = v^N G_N.
\end{equation}
Now for the channel splitting step we have to distinguish whether we want to
decode a bit from sender~1 or from sender~2 as follows 
\begin{equation}
W_N^{(b_k,i,j)}= 
\begin{cases}
W_N^{(0,i,j)}: u_i \rightarrow \rho^{U^{i-1}_1 V_1^j B^N}_{(0,i,j),u_i}
\quad \quad \mathrm{if} \; b_k =0 \\ 
W_N^{(1,i,j)}: v_j \rightarrow \rho^{U^{i}_1 V_1^{j-1} B^N}_{(1,i,j),v_j}
\quad \quad \mathrm{if} \; b_k =1%
\end{cases}%
\end{equation}
with output states 
\begin{align}
\rho^{U^{i-1}_1 V_1^j B^N}_{(0,i,j),u_i}=\sum_{u_1^{i-1}, v_1^j}&\frac{1}{2^{k-1}}\ketbra{}{u_1^{i-1}}{u_1^{i-1}} \nonumber\\
&\otimes \ketbra{}{v_1^j}{v_1^j}\otimes \bar\rho^{B^N}_{u_1^i,v_1^j}, \\
\rho^{U^i_1 V_1^{j-1} B^N}_{(1,i,j),v_j}=\sum_{u_1^{i}, v_1^{j-1}}&\frac{1}{2^{k-1}}\ketbra{}{u_1^{i}}{u_1^{i}}\nonumber\\
&\otimes \ketbra{}{ v_1^{j-1}}{v_1^{j-1}} \otimes \bar\rho^{B^N}_{u_1^i,v_1^j},
\end{align}
where
\begin{equation}
\bar\rho^{B^N}_{u_1^i,v_1^j} = \sum_{u_{i+1}^N,v_{j+1}^N} \frac{1}{2^{2N-k}}
\rho^{B^N}_{u^N,v^N}.
\end{equation}
Similar to the case of classical-quantum polar coding for a single sender,
we now discuss how a \textit{quantum successive cancellation decoder} operates for the
cq-MAC. As in \cite{WG13} we can build projectors to decide whether the $k$th input, corresponding to the split channel $W_N^{(b_k,i,j)}$, is equal to zero or
one: 
\begin{equation}
\Pi_{(b_k,i,j),0}= 
\begin{cases}
\Pi^{U^{i-1}_1 V_1^j B^N}_{(0,i,j),0} \quad \quad \mathrm{if} \; b_k =0 \\ 
\Pi^{U^i_1 V_1^{j-1} B^N}_{(1,i,j),0} \quad \quad \mathrm{if} \; b_k =1%
\end{cases}%
\end{equation}
with 
\begin{align}
\Pi^{U^{i-1}_1 V_1^j B^N}_{(0,i,j),0} = \left\{ \sqrt{\rho^{U^{i-1}_1 V_1^j
B^N}_{(0,i,j),0}} - \sqrt{\rho^{U^{i-1}_1 V_1^j B^N}_{(0,i,j),1}} \geq 0 \right\}
\\
\Pi^{U^i_1 V_1^{j-1} B^N}_{(1,i,j),0} = \left\{ \sqrt{\rho^{U^i_1 V_1^{j-1}
B^N}_{(1,i,j),0}} - \sqrt{\rho^{U^i_1 V_1^{j-1} B^N}_{(1,i,j),1}} \geq 0 \right\}
\end{align}
and 
\begin{equation}
\Pi_{(b_k,i,j),1} = \id - \Pi_{(b_k,i,j),0}.
\end{equation}
The notation $\{A\geq 0\}$  denotes the projector onto the positive eigenspace of $A$, and
 $\{A < 0\}$ denotes the projector onto its negative eigenspace.

Note that, similar to \cite{WG13}, we can write 
\begin{align}
\Pi^{U^{i-1}_1 V_1^j B^N}_{(0,i,j),0} = \sum_{u_1^{i-1}, v_1^j} &\ketbra{}{u_1^{i-1}}{u_1^{i-1}}\otimes \ketbra{}{v_1^j}{v_1^j} \nonumber\\
&\otimes \Pi^{B^N}_{(0,i,j),u_1^{i-1} 0,v_1^j} , \\
\Pi^{U^i_1 V_1^{j-1} B^N}_{(1,i,j),0} = \sum_{u_1^{i}, v_1^{j-1}} &\ketbra{}{u_1^{i}}{u_1^{i}}\otimes \ketbra{}{ v_1^{j-1}}{v_1^{j-1}}  \nonumber\\
&\otimes \Pi^{B^N}_{(1,i,j),u_1^i, v_1^{j-1}0},
\end{align}
where
\begin{align}
\Pi^{B^N}_{(0,i,j),u_1^{i-1}0, v_1^j } =& \left\{ \sqrt{\bar\rho^{B^N}_{u_1^{i-1}0,v_1^j}} - \sqrt{\bar\rho^{B^N}_{u_1^{i-1}1,v_1^j}} \geq 0 \right\} ,
\\
\Pi^{B^N}_{(1,i,j),u_1^i, v_1^{j-1}0} =& \left\{ \sqrt{\bar\rho^{B^N}_{u_1^i,v_1^{j-1}0}} - \sqrt{\bar\rho^{B^N}_{u_1^i,v_1^{j-1}1}} \geq 0 \right\}.
\end{align}

Now again with arguments presented in \cite{WG13} we get a POVM with
elements 
\begin{multline}
\Lambda_{u^N,v^N}  = \Pi_{(b_1,i_1,j_1),\{u_1, v_1\} } \\
\cdots \Pi_{(b_{2N},i_{2N},j_{2N}),\{u_1^{N-1}u_Nv_1^N, u_1^Nv_1^{N-1}v_N\} } \\
\cdots \Pi_{(b_1,i_1,j_1),\{u_1,v_1\} } ,
\end{multline}
where the exact values of $i$ and $j$ depend on the monotone chain rule
chosen for decoding, as well as whether a projector attempts to
decode $u_i$ or $v_j$. As required for a POVM we also have that
\begin{equation}
\sum_{u_A, v_A} \Lambda_{u^N,v^N} = \id^{B^N},
\end{equation}
by noting that we can set $\Pi_{(b_k,i,j),\{u_i, v_j\} } = \id$ when $\{u_i,
v_j\}$ is a frozen bit.

Using the bitwise projections we can build the successive cancellation
decoder with the decoding rules: 
\begin{align}
\hat u_i &= 
\begin{cases}
u_i \quad \quad & \mathrm{if} \; i\in A^c \\ 
h(\hat u_1^{i-1}, \hat v_1^j) \quad \quad & \mathrm{if} \; i\in A%
\end{cases} ,
\\
\hat v_j &= 
\begin{cases}
v_j \quad \quad & \mathrm{if} \; j\in A^c \\ 
g(\hat u_1^{i}, \hat v_1^{j-1}) \quad \quad & \mathrm{if} \; j\in A%
\end{cases}%
,
\end{align}
where $h(\hat u_1^{i-1}, \hat v_1^j)$ is the outcome of the $k$th
measurement when $b_k=0$ based on 
\begin{equation}
\{ \Pi^{B^N}_{(0,i,j),u_1^{i-1}0, v_1^j } , \Pi^{B^N}_{(0,i,j),u_1^{i-1}1, v_1^j } \},
\end{equation}
and $g(\hat u_1^{i}, \hat v_1^{j-1})$ is the outcome when $b_k=1$ based on 
\begin{equation}
\{ \Pi^{B^N}_{(1,i,j),u_1^i, v_1^{j-1}0} , \Pi^{B^N}_{(1,i,j),u_1^i, v_1^{j-1}1} \}.
\end{equation}

Due to the structure of the decoder and the polarization effect,
the block error probability
decays exponentially with the number of channel uses as in the single-sender case described in the previous
section.

\subsection{Continuity of rates and approximations}\label{rates}

We now argue that the above approach can be used to achieve
the entire dominant face of the rate region. This task was achieved for the
classical Slepian-Wolf source coding problem \cite{A12} involving rates
based on the conditional Shannon entropy, and then extended to the complementary problem
of channel coding over the classical MAC \cite{O13}. We now extend that
technique by applying it to channel coding
for the classical-quantum MAC.

We start by defining a distance measure. Let $b^{2N},\tilde b^{2N}$ denote
two paths and $(R_u, R_v), (\tilde R_u, \tilde R_v)$ their corresponding
rate pairs. Then we define the distance between the two paths $b^{2N}$ and $%
\tilde b^{2N}$ as follows 
\begin{equation}
d(b^{2N},\tilde b^{2N}) \equiv |R_u - \tilde R_u | = |R_v - \tilde R_v | ,
\end{equation}
where the last equality holds since $R_u +R_v = \tilde R_u +\tilde R_v =
I(XY;B)$.

We now define two paths $b^{2N},\tilde b^{2N}$ to be neighbors if $\tilde
b^{2N}$ can be obtained from $b^{2N}$ by transposing $b_i$ with $b_j$ for
some $i<j$ such that $b_i\neq b_j$ and $b_{i-1}^{j+1}$ is either all 0 or
all 1. We state the following proposition, which bounds the distance between two
neighboring paths. This generalizes Proposition~3 in \cite{A12} to the
quantum setting considered here. 
\begin{prop}\label{neig}
If paths $b^{2N}$ and $\tilde b^{2N}$ are neighboring, then the following holds 
\begin{equation}
d(b^{2N},\tilde b^{2N}) \leq \frac{1}{N}.
\end{equation}
\end{prop}
\begin{IEEEproof}
Let $b^{2N}$ be a path with edge variables $S^{2N}$ and let $\tilde b^{2N}$ differ from $b^{2N}$ by a transposition in the coordinates $i<j$. 

First we check the case, where $b_i=0$ and $b_j=1$ and the string $b_{i+1}^{j-1}$ contains only 1's. 
The rate difference can be written as
\begin{multline}
R_u - \tilde R_u = \frac{1}{N} [ I( S_i ; B^N| S^{i-1}) \\
-I(S_i ; B^N | S^{i-1},S_j,S_{i+1}^{j-1}) ] ,
\end{multline}
and we can see directly that
\begin{equation}
R_u - \tilde R_u \leq \frac{1}{N},
\end{equation}
where the inequality follows from the non-negativity of the conditional mutual information and a dimension bound when one unconditioned system is classical. 

Since $R_u - \tilde R_u \geq 0$ we can conclude
\begin{equation}
\abs{R_u - \tilde R_u} \leq \frac{1}{N} .
\end{equation}
This holds for $b_i^j = 01^{j-i}$. The three other options $( b_i^j \in \{ 0^{j-i}1, 1^{j-i}0, 10^{j-i} \} )$ can be proven similarly, by exchanging the roles of  $b^{2N}$ and $\tilde b^{2N}$ or by considering $R_v - \tilde R_v$ or both. 
\end{IEEEproof}
Using this result we can show that the distribution of achievable points on
the dominant face of the rate region is dense, and we state the following
theorem generalizing Theorem 1 in \cite{A12} to the quantum case considered here. 
\begin{thm}\label{thm}
Let $(R_x, R_y)$ be a given rate pair on the dominant face. For any given $\epsilon >0$, there exists an $N \in \mathbb{N}$ and a chain rule $b^{2N}$ on $U^NV^N$ such that $b^{2N}$ belongs to the class $\nu_{2N}  = \{ 0^i 1^N 0^{N-i} : 0 \leq i \leq N \}$ and has a rate pair $(R_1, R_2)$ satisfying 
\begin{equation}
|R_1 - R_x| \leq \epsilon \qquad\textit{and}\qquad |R_2 - R_y| \leq \epsilon.
\end{equation}
\end{thm}
The proof follows from the fact that in $\nu_{2N}$ two paths $0^i 1^N 0^{N-i}
$ and $0^{i+1} 1^N 0^{N-i-1}$ are neighbors, thus we simply have to fix $N > 
\frac{1}{\epsilon}$ and Theorem \ref{thm} follows from Proposition \ref{neig}.

Note that this result is only concerned with permutations from the class $%
\nu_{2N}$. This is sufficient for our purposes, but we note that it may be
interesting to consider a more general class when choosing the decoding path.

\subsection{Path scaling and polarization}\label{scaling}

It was shown in the previous section that one can find polar codes that
approximate points on the dominant face of the rate region for a cq-MAC.
Here we show that these approximations are stable under scaling of the
chosen path, implying that performing a step of the polar coding
recursion does not change the achievable rates so long as polarization still
holds. We generalize ideas from \cite{A12} to the classical-quantum MAC.

We will look at paths $kb^{2N}$ which denote the scaling of a path $b^{2N}$
as 
\begin{equation}
\underbrace{b_{1}\dots b_{1}}_{k}\underbrace{b_{2}\dots b_{2}}_{k}\dots
\dots \underbrace{b_{2N}\dots b_{2N}}_{k}
\end{equation}%
and therefore represent a monotone chain rule for $U^{kN}V^{kN}$. Note that
we can write a step of the polar-code transformation as 
\begin{equation}
T_{2i-1}=S_{i}\oplus \tilde{S}_{i},\ \ \ \ T_{2i}=S_{i},
\end{equation}%
and thus we can show that an additional step of polarization does not affect
the rate 
\begin{align}
&\!\!\!\!\!\! I(T_{2i-1};B^{N}|T^{2i-2})+I(T_{2i};B^{N}|T^{2i-1}) \nonumber\\
& =I(T_{2i-1}T_{2i};B^{N}|T^{2i-2}) \nonumber\\
& =I(S_{i}\oplus \tilde{S}_{i},S_{i};B^{N}|S^{i-1}\oplus \tilde{S}^{i-1},S^{i-1}) \nonumber\\
& =I(\tilde{S}_{i},S_{i};B^{N}|\tilde{S}^{i-1},S^{i-1}) \nonumber\\
& =2I(S_{i};B^{N}|S^{i-1}) ,
\end{align}
where the last step follows if $T^{4N}$ follows the path $2b^{2N}$. From
this we can conclude that if a path $b^{2N}$ achieves a rate $(R_{1},R_{2})$
then the path $2b^{2N}$ achieves the same rate pair.

Now the polarization argument follows directly using arguments from the
single-sender case in~\cite{WG13}.

\subsection{Polar code performance}

Since we use a POVM with the same basic structure as the single-sender case,
the analysis of the error probability follows the same arguments.
That is, by applying the non-commutative union bound from \cite{S11}
to the probability of error $P_e(M, b^{2N}, (K_u,K_v), (A_u, A_v), (u_{A_u^c}, v_{A_v^c}))$ for code length $M$ with $M=2^mN$ and $m$ a positive integer, a chosen path $b^{2N}$, the number of information bits $(K_u,K_v)$ and sets of information bits $(A_u, A_v)$ for each sender and the choice for the frozen bits $(u_{A_u^c}, v_{A_v^c})$,
we get 
\begin{multline}
P_e(M, b^{2N}, (K_u,K_v), (A_u, A_v), (u_{A_u^c}, v_{A_v^c}))  \\
 \leq 2 \sqrt{ \sum_{i\in A_u, j\in A_v} \frac{1}{2} \sqrt{ F(W_N^{(b_k,i,j)})}}.
\end{multline}
Therefore, we can state that the error probability in \cite{WG13} (see Section \ref{cq}) also
holds for multiple-user settings 
\begin{equation}
P_e(M, R) = o(2^{-\frac{1}{2}M^\beta}),
\end{equation}
with $M=2^mN$.

\section{Three sender MAC}

\label{3MAC} We can easily extend the approach for two senders discussed above to the
case of many senders. For our purposes, we are particularly interested in the
three-sender setting. Therefore, similar to the two-dimensional case, we simply follow a path through a
three-dimensional cube (see Figure \ref{3dpath} for example),
in order to choose a path $b_k\in \{0,1,2\}$ giving rise to the following
achievable rates 
\begin{align}
\begin{split}
R_x = \frac{1}{N} \sum_{k:b_k=0} I(S_k;B^N|S^{k-1}) \leq I(X;B|YZ), \\
R_y = \frac{1}{N} \sum_{k:b_k=1} I(S_k;B^N|S^{k-1}) \leq I(Y;B|XZ), \\
R_z = \frac{1}{N} \sum_{k:b_k=2} I(S_k;B^N|S^{k-1}) \leq I(Z;B|XY), \\
R_x + R_y \leq I(XY;B|Z), \\
R_x + R_z \leq I(XZ;B|Y), \\
R_y + R_z \leq I(ZY;B|X), \\
R_x + R_y + R_z = I(XYZ;B).
\end{split}
\end{align}

\newcommand\XA{0.5}
\newcommand\Thick{3pt}

\begin{figure}
\centering
\begin{tikzpicture}[scale=0.95]
\foreach \z in {0,...,5}{
 \foreach \y in {0,...,5}{
  \foreach \x in {0,...,5}{
   	 \draw[ultra thin] (\x-\XA*\y,0+\XA*\y) -- (\x-\XA*\y,5+\XA*\y);
 	 \draw[ultra thin] (0-\XA*\y,\x+\XA*\y) -- (5-\XA*\y,\x+\XA*\y);
	 \draw[ultra thin] (0+\y,0+\z) -- (-5*\XA+\y,5*\XA+\z);
   }
  }
}
   	 \draw[line width=\Thick] (0,0) -- (0,5);
	\draw[line width=\Thick] (0,0) -- (5,0);
	\draw[line width=\Thick] (5,5) -- (0,5);
	\draw[line width=\Thick] (5,5) -- (5,0);
	\draw[line width=\Thick] (5,5) -- (2.5,7.5);
	\draw[line width=\Thick] (-2.5,2.5) -- (-2.5,7.5);
	\draw[line width=\Thick] (2.5,7.5) -- (-2.5,7.5);
	\draw[line width=\Thick] (2.5,7.5) -- (2.5,2.5);
	\draw[line width=\Thick] (0,5) -- (-2.5,7.5);
	\draw[line width=\Thick] (-2.5,2.5) -- (0,0);
	\draw[line width=\Thick] (5,0) -- (2.5,2.5);
	\draw[line width=\Thick] (2.5,2.5) -- (-2.5,2.5);
  \foreach \x in {1,...,5}{
	\draw (-2.7 + \x*0.5,2.3 - \x*0.5)node  {$u_{\x}$};
}
  \foreach \x in {1,...,5}{
	\draw (-2.7,7.5 - \x)node  {$v_{\x}$};
}
  \foreach \x in {1,...,5}{
	\draw (-2.7 + \x,7.7)node  {$w_{\x}$};
}
\draw (-2.7 ,7.7)node  {$\emptyset$};
\draw (5.6,-0.2)node  {$u_5v_5w_5$};
\draw[line width=3.5pt, black!40] (-2.5, 7.5) -- (-0.5, 7.5) -- (0.5, 6.5) -- (0.5, 1.5) -- (2, 0) -- (5, 0);
	\draw[line width=\Thick] (5,5) -- (0,5);
	\draw[line width=\Thick] (2.5,2.5) -- (-2.5,2.5);
\end{tikzpicture}
\caption{Order for decoding the 3-user MAC, with example path in grey. Allowed paths are only those which go only to the right, front or bottom at each step, to imply a monotone chain rule. This is important to ensure the correct decoding order for the successive cancellation decoder.}\label{3dpath}
\end{figure}
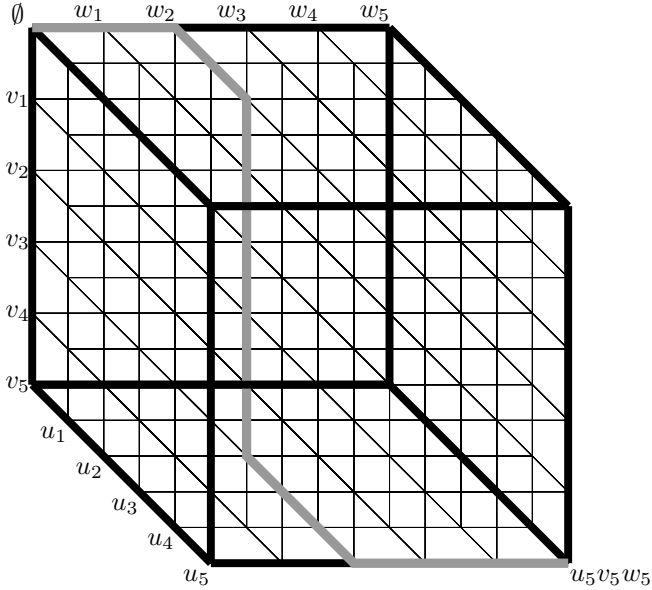

Note that for the $m$-sender case we can generalize the above coding method
by simply following a path in an $m$-dimensional structure while making sure
that the entropy equations remain monotonic.

\section{Universal polar codes for the compound MAC}\label{comMAC}

\label{cMAC}

Next we will describe how the so-called ``universal polar codes'' introduced
in \cite{HU13} can be applied to the cq-MAC to achieve rates
for compound channels. In particular we will make use of the second scheme
described in \cite{HU13} and the generalizations of this scheme to MACs in 
\cite{WS14}.

For now we will look at compound MACs based on sets of two different MACs. 
The essential approach is to \textquotedblleft align\textquotedblright\
polarized indices as follows. Note that we assume that the selected channel
is known to the receiver but not to the sender. For simplicity we consider two
2-sender MACs with equal sum rate. It is clear that a standard polar code
which is good for one of the channels is not necessarily good for the other
one. To get around this issue we align the two senders independently. 

Conceptually the approach is similar to the classical setting, but needs to be applied to the sets of good and bad classical-quantum channels, defined by upper bounding the fidelity of the polarized channels. 

Recall that in Section \ref{cq} we reviewed the channel splitting step for classical-quantum channel polarization, first introduced in \cite{WG13}. Similarly, here we define the partial split channels $P_{i}:U_{i}\rightarrow
B^{N}U_{1}^{i-1}V_{1}^{j}$ and $Q_{i}:U_{i}\rightarrow
B^{N}U_{1}^{i-1}V_{1}^{j}$, each corresponding to the first
sender of one of the two MACs in the set comprising the compound MAC. In the previous section, $b_k$ served as a label indicating which sender should be decoded in the $k$th step.
Moreover, the channels $P_i$ and $Q_i$ can be considered to be equivalent to looking at only the channel uses of the corresponding MACs for which $b_k=0$. 
Let 
\begin{equation}
\begin{split}
\mathcal{G}_{(1)}& =\left\{i\in \lbrack 1:N]:\sqrt{F(P_{i})}<2^{-N^{\beta }}\right\}, \\
\mathcal{G}_{(2)}& =\left\{i\in \lbrack 1:N]:\sqrt{F(Q_{i})}<2^{-N^{\beta }}\right\}, \\
\mathcal{B}_{(1)}& =\left\{i\in \lbrack 1:N]:\sqrt{F(P_{i})}\geq 2^{-N^{\beta }}\right\}, \\
\mathcal{B}_{(2)}& =\left\{i\in \lbrack 1:N]:\sqrt{F(Q_{i})}\geq 2^{-N^{\beta }}\right\}.
\end{split}%
\end{equation}%
denote the sets of indices corresponding to whether a bit is good or bad for a
channel. These sets tell us whether the attempt of sending an information
bit through one of the MACs would be successful with high probability for
the $i$th channel use of the first sender. Due to the polarization
effect, all bits will be in one of the following sets: 
\begin{align}
\mathcal{A}_{\text{I}}\ & =\mathcal{G}_{(1)}\cap \mathcal{G}_{(2)}, \\
\mathcal{A}_{\text{II}}\ & =\mathcal{G}_{(1)}\cap \mathcal{B}_{(2)}, \\
\mathcal{A}_{\text{III}}& =\mathcal{B}_{(1)}\cap \mathcal{G}_{(2)}, \\
\mathcal{A}_{\text{IV}}& =\mathcal{B}_{(1)}\cap \mathcal{B}_{(2)}.
\end{align}%
Bits belonging to the sets $\mathcal{A}_{\text{I}}$ will also be decoded
with high probability in the compound setting and bits in $\mathcal{A}_{%
\text{IV}}$ will have to be set as frozen bits. Due to Theorem \ref{thm} in Section \ref{rates} we can find monotone chain rules for each MAC which approximate every point on the dominant face of the rate region of the corresponding MAC.

The main idea from here is to align the sets $\mathcal{A}_\text{II}$ and $\mathcal{A}_%
\text{III}$ within a recursion to achieve the capacity of the compound MAC. We will do so alternating in each step of
the recursion either for the first or the second sender. Here, just as in
the classical case, we have to ensure that we align bits such that the
successive cancellation decoder can still be applied.

We take two polar coding blocks which have both already been
polarized independently of each other. Since both blocks have been built
from the same channel, the sets of indices are identical for both blocks. We
then combine the first index from $\mathcal{A}_\text{II}$ in the first block
with the first index of $\mathcal{A}_\text{III}$ in the second block by an
additional CNOT gate, and similarly for the second indices and so on. With such a scheme,
we can halve the fraction of incompatible indices, those from
the sets $\mathcal{A}_\text{II}$ and $\mathcal{A}_\text{III}$, for the first
sender.

Intuitively this can be seen as sending the same information bits via both
of the aligned channels, so that the reciever will be able to decode one of
them independently of which MAC is actually used. Since we assume that the
reciever knows the used channel, this works well with the successive
cancellation decoder, because the reciever can just decode the channel which
is good for the used MAC first and then decode the aligned channel as if it
is a frozen bit.

In the next step we take two of the blocks after the first iteration step
and repeat the process for the second sender. Hence, we again halve the
fraction of incompatible indices for this sender. In the following we
repeat this process until the fraction of incompatible indices tends to
zero.

\pgfooclass{stamp}{     \method stamp() {     }
 \method cnot(#1,#2,#3,#4) {         	\draw (#1,#2) -- (#1-#3,#2) -- (#1-#3,#2-#4) -- (#1,#2-#4);
	\draw (#1-#3,#2) circle (0.1);
	\draw (#1-#3,#2+0.1) -- (#1-#3,#2) -- (#1-#3-0.25,#2);
	\draw[dotted] (#1-#3-0.25,#2) -- (#1-#3-0.75,#2);
   }
} \pgfoonew \mystamp=new stamp()

\begin{figure}
\centering
\begin{tikzpicture}[scale=1]
	\draw (0,0) -- (1,0) -- (1,1) -- (0,1) -- cycle; 
	\node[font=\tiny] at (0.5,0.5) {$U_{2b}$};
	\draw (0,1) -- (1,1) -- (1,3) -- (0,3) -- cycle; 
	\node[font=\tiny] at (0.5,2) {$V_2$};
	\draw (0,3) -- (1,3) -- (1,4) -- (0,4) -- cycle; 
	\node[font=\tiny] at (0.5,3.5) {$U_{2a}$};
	\draw (0,5) -- (1,5) -- (1,6) -- (0,6) -- cycle; 
	\node[font=\tiny] at (0.5,5.5) {$U_{1b}$};
	\draw (0,6) -- (1,6) -- (1,8) -- (0,8) -- cycle; 
	\node[font=\tiny] at (0.5,7) {$V_1$};
	\draw (0,8) -- (1,8) -- (1,9) -- (0,9) -- cycle; 
	\node[font=\tiny] at (0.5,8.5) {$U_{1a}$};
	\mystamp.cnot(0,8.7,1,8);
	\mystamp.cnot(0,8.4,1.3,8);
           \draw[->,solid] (-0.5,9) node[font=\tiny, below right=0]{1} -- (-0.5,8.7); 			
	\draw[->,densely dotted] (-0.5,8.7)node[font=\tiny, below right=0]{2} -- (-0.5,8.4);
	\draw[->] (-0.5,8.4)node[font=\tiny, below right=0]{3} -- (-0.5,5);
          \draw[->] (-0.5,4)node[font=\tiny, below right=0]{1} -- (-0.5,0.7);			 
	\draw[->,densely dotted] (-0.5,0.7)node[font=\tiny, below right=0]{2} -- (-0.5,0.4);
	\draw[->] (-0.5,0.4)node[font=\tiny, below right=0]{3} -- (-0.5,0);
\end{tikzpicture}
\caption{Coding for compound MAC: first iteration step. The arrows indicate
the decoding order and must be followed in ascending order of the attached
numbers.}
\label{decMAC}
\end{figure}
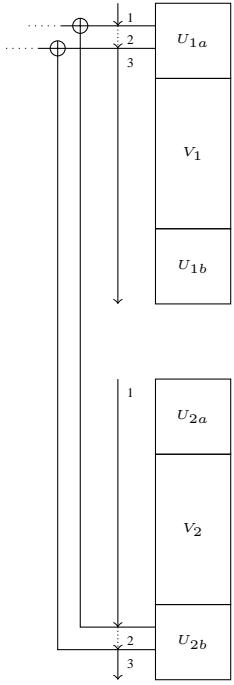

To generalize the above scheme to the $k$-user MAC we simply decode by
alternating over the different senders in each step of the recursion.
Therefore each sender becomes aligned in every $k$-th step of the recursion.
In the recursion $mK$ steps reduce the fraction of incompatible indices for
the $K$-user MAC to $(|\mathcal{A}_\text{II}| + |\mathcal{A}_\text{III}|)/2^mN$%
. For example for the case of two users this means that every two steps we
halve the fraction of incompatible indices. This is due to the fact that we
need one recursion step for each sender to halve the fraction of incompatible
indices for that particular sender. Then we can reorder our decoding in a
way that the successive cancellation decoder still holds. Figure \ref{decMAC}
illustrates the process.

In order to use the compound MAC for the interference channel we need to generalize the described approach to the setting of unequal sum rates. 
Therefore assume that we want to code for a rate pair $(R_x,R_y)$ on the dominant face of the achievable rate region for the compound MAC consisting of a set of two MACs. 
Now we can find a rate pair  $(R'_x,R'_y)$ for the first MAC in that set and a rate pair $(R''_x,R''_y)$ for the second MAC such that 
\begin{align}
R_x &\leq \min(R'_x,R''_x) \\
R_y &\leq \min(R'_y,R''_y).
\end{align}
We can use the corresponding monotone chain rules for these two MACs to code for the targeted point on the rate region of the compound MAC.
Note that in the setting of unequal sum-rate the sets $\mathcal{A}_\text{II}$ and $\mathcal{A}_\text{III}$ are not necessarily of equal size.
This is not a problem for the aligning process, because we can simply align until one of the sets has no unaligned indices left and then handle the remaining indices in the larger set as frozen indices. It is easy to see that this is sufficient to code at rates $\min(R'_x,R''_x)$ and $\min(R'_y,R''_y)$ and therefore achieve the dominant face for the achievable rate region of the compound MAC.

It was previously unknown how to code for the $k$-user compound MAC in a
quantum setting. Having this result will allow us to also code for
classical-quantum interference channels, as discussed in the next section.

\section{Interference channel}

\label{int}

The two-user classical-quantum interference channel \cite{SFWSH11, FHSSW12},
as discussed in Section \ref{cqMulti}, can be represented by
its set of output states as follows:
\begin{equation}
\{\rho _{x_{1},x_{2}}^{B_{1}B_{2}}\}_{x_{1}\in \mathcal{X}_{1},x_{2}\in 
\mathcal{X}_{2}}.
\end{equation}
In the classical setting the best known achievable rate region for the
interference channel is given by the Han-Kobayashi rate region \cite%
{HK81}. We now show that this region can be achieved for the two-user
classical-quantum interference channel by using polar codes. The scheme is a
direct generalization of that presented in \cite{WS14}. Indeed note that the conceptual approach is independent of  whether we apply it to classical or classical-quantum channels. Therefore the
Han-Kobayashi rate region can be achieved by splitting the message of the
first sender $m_{1}$ into two parts labelled $(l_{1},l_{2})$ and similarly
for the second sender the message $m_{2}$ is split into $(l_{3},l_{4})$. Now
we get channel inputs represented by the random variables $X_{1}^{N}$ and $%
X_{2}^{N}$ via symbol-to-symbol encoding maps $x_{1}(v_{1},v_{2})$ and $%
x_{2}(v_{3},v_{4})$ corresponding to the codewords $v_{j}^{N}(L_{j})$. Then
receiver 1 decodes from $\rho _{x_{1},x_{2}}^{B_{1}}$ the messages $%
(l_{1},l_{2},l_{3})$ and receiver 2 decodes $\rho _{x_{1},x_{2}}^{B_{2}}$ to
get the message triple $(l_{2},l_{3},l_{4})$. The Han-Kobayashi rate region
is defined for rates $(S_{1},S_{2},T_{1},T_{2})$ as follows:
\begin{equation}
\begin{split}
S_{1}& \leq I(V_{1};B_{1}|V_{3}V_{4}), \\
T_{1}& \leq I(V_{3};B_{1}|V_{1}V_{4}), \\
T_{2}& \leq I(V_{4};B_{1}|V_{1}V_{3}), \\
S_{1}+T_{1}& \leq I(V_{1}V_{3};B_{1}|V_{4}), \\
S_{1}+T_{2}& \leq I(V_{1}V_{4};B_{1}|V_{3}), \\
T_{1}+T_{2}& \leq I(V_{3}V_{4};B_{1}|V_{1}), \\
S_{1}+T_{1}+T_{2}& \leq I(V_{1}V_{3}V_{4};B_{1}),
\end{split}
\end{equation}
\begin{equation}
\begin{split}
S_{2}& \leq I(V_{2};B_{2}|V_{3}V_{4}), \\
T_{1}& \leq I(V_{3};B_{2}|V_{2}V_{4}), \\
T_{2}& \leq I(V_{4};B_{2}|V_{2}V_{4}), \\
S_{2}+T_{1}& \leq I(V_{2}V_{3};B_{2}|V_{4}), \\
S_{2}+T_{2}& \leq I(V_{2}V_{4};B_{2}|V_{3}), \\
T_{1}+T_{2}& \leq I(V_{3}V_{4};B_{2}|V_{2}), \\
S_{2}+T_{1}+T_{2}& \leq I(V_{2}V_{3}V_{4};B_{2}),
\end{split}
\end{equation}%
The achievable rate region for the interference channel is the set of all
rates $(S_{1}+T_{1},S_{2}+T_{2})$.

This can be seen as an interference network with four senders $x_1(v_1, v_2)$
and $x_2(v_3, v_4)$ and two recievers. The rate tuples $(S_1, T_1, T_2)$ and 
$(S_2, T_1, T_2)$ coincide with two 3-user MAC regions with two common
senders and the intersection of both gives the rate region $(S_1, S_2, T_1,
T_2)$. We can find monotone chain rules for every point on the dominant
face of each MAC. Now we use the approach for the compound MAC to align the
two common senders along with the results of the previous section in
order to achieve the rate region for the two-user interference channel using
polar coding. Note that it is necessary to use the approach presented for
compound MACs in order to ensure that the decoding used for the polar codes
is good for both MACs. With this approach we can achieve the Han-Kobayashi
rate region by a successive cancellation decoder for each receiver.
There are known techniques for handling arbitrary input distributions. Rather than
go into detail, we point to the work in \cite{WS14} which elaborates on this point.

\section{Conclusion}

\label{concl} We have applied the recently introduced polar coding technique
to achieve known rates for a variety of classical-quantum multi-user
channels, with our main result being that the Han-Kobayashi rate region for the
two-user interference channel can be achieved by a successive cancellation
decoder via polar coding. In particular, we emphasize that this was achieved
without the use of a quantum simultaneous decoder. The interference channel model
forms a basis from which other multi-user channels can be built. This result
and the wide range of problems for which polar coding has been applied in
classical information theory suggest that it might be possible to generalize a wide range of
problems to the classical-quantum setting using a successive decoder and in
particular without the need of a quantum simultaneous decoder. 

\bibliographystyle{IEEEtran}
\bibliography{IEEEabrv,Bib}

\begin{thebibliography}{10}
\providecommand{\url}[1]{#1}
\csname url@samestyle\endcsname
\providecommand{\newblock}{\relax}
\providecommand{\bibinfo}[2]{#2}
\providecommand{\BIBentrySTDinterwordspacing}{\spaceskip=0pt\relax}
\providecommand{\BIBentryALTinterwordstretchfactor}{4}
\providecommand{\BIBentryALTinterwordspacing}{\spaceskip=\fontdimen2\font plus
\BIBentryALTinterwordstretchfactor\fontdimen3\font minus
  \fontdimen4\font\relax}
\providecommand{\BIBforeignlanguage}[2]{{%
\expandafter\ifx\csname l@#1\endcsname\relax
\typeout{** WARNING: IEEEtran.bst: No hyphenation pattern has been}%
\typeout{** loaded for the language `#1'. Using the pattern for}%
\typeout{** the default language instead.}%
\else
\language=\csname l@#1\endcsname
\fi
#2}}
\providecommand{\BIBdecl}{\relax}
\BIBdecl

\bibitem{W01}
A.~Winter, ``The capacity of the quantum multiple access channel,'' \emph{IEEE
  Transactions on Information Theory}, vol.~47, no.~7, pp. 3059--3065, November
  2001, arXiv:quant-ph/9807019.

\bibitem{W99}
------, ``Coding theorem and strong converse for quantum channels,'' \emph{IEEE
  Transactions on Information Theory}, vol.~45, no.~7, pp. 2481--2485, November
  1999, arXiv:1409.2536.

\bibitem{FHSSW12}
O.~Fawzi, P.~Hayden, I.~Savov, P.~Sen, and M.~M. Wilde, ``Classical
  communication over a quantum interference channel,'' \emph{IEEE Transactions
  on Information Theory}, vol.~58, no.~6, pp. 3670--3691, June 2012,
  arXiv:1102.2624.

\bibitem{Carleial75}
A.~B. Carleial, ``A case where interference does not reduce capacity,''
  \emph{IEEE Transactions on Information Theory}, vol.~21, no.~5, pp. 569--570,
  September 1975.

\bibitem{HK81}
T.~S. Han and K.~Kobayashi, ``A new achievable rate region for the interference
  channel,'' \emph{IEEE Transactions on Information Theory}, vol.~27, no.~1,
  pp. 49--60, January 1981.

\bibitem{SFWSH11}
I.~Savov, O.~Fawzi, M.~M. Wilde, P.~Sen, and P.~Hayden, ``Quantum interference
  channels,'' \emph{Proceedings of the 49th Annual Allerton Conference on
  Communication, Control, and Computing}, pp. 609--616, September 2011,
  arXiv:1102.2955.

\bibitem{S11}
P.~Sen, ``Achieving the {Han-Kobayashi} inner bound for the quantum
  interference channel by sequential decoding,'' \emph{Proceedings of the 2012
  IEEE International Symposium on Information Theory}, pp. 736--740, July 2012,
  arXiv:1109.0802.

\bibitem{FS12}
O.~Fawzi and I.~Savov, ``Rate-splitting in the presence of multiple
  receivers,'' July 2012, arXiv:1207.0543.

\bibitem{A09}
E.~Arikan, ``Channel polarization: A method for constructing capacity-achieving
  codes for symmetric binary-input memoryless channels,'' \emph{IEEE
  Transactions on Information Theory}, vol.~55, no.~7, pp. 3051--3073, July
  2009, arXiv:0807.3917.

\bibitem{STY10}
E.~\c{S}a\c{s}o\u{g}lu, E.~Telatar, and E.~M. Yeh, ``Polar codes for the
  two-user binary-input multiple-access channel,'' \emph{2010 IEEE Information
  Theory Workshop}, pp. 1--5, January 2010.

\bibitem{STY13}
------, ``Polar codes for the two-user multiple-access channel,'' \emph{IEEE
  Transactions on Information Theory}, vol.~59, no.~10, pp. 6583--6592, October
  2013, arXiv:1006.4255.

\bibitem{O13}
A.~\"Onay, ``Successive cancellation decoding of polar codes for the two-user
  binary-input {MAC},'' \emph{Proceedings of the 2013 IEEE International
  Symposium on Information Theory}, pp. 1122--1126, July 2013.

\bibitem{MELK13}
H.~Mahdavifar, M.~El-Khamy, J.~Lee, and I.~Kang, ``Achieving the uniform rate
  region of multiple access channels using polar codes,'' July 2013,
  arXiv:1307.2889.

\bibitem{GAG13}
N.~Goela, E.~Abbe, and M.~Gastpar, ``Polar codes for broadcast channels,''
  \emph{Proceedings of the 2013 IEEE International Symposium on Information
  Theory}, pp. 1127--1131, 2013, arXiv:1301.6150.

\bibitem{MHSU14}
M.~Mondelli, S.~H. Hassani, I.~Sason, and R.~Urbanke, ``Achieving {Marton's}
  region for broadcast channels using polar codes,'' \emph{IEEE Transactions on
  Information Theory}, vol.~61, no.~2, pp. 783--800, February 2015,
  arXiv:1401.6060.

\bibitem{WS14}
L.~Wang and E.~\c{S}a\c{s}o\u{g}lu, ``Polar coding for interference networks,''
  January 2014, arXiv:1401.7293.

\bibitem{A12}
E.~Arikan, ``Polar coding for the {Slepian-Wolf} problem based on monotone
  chain rules,'' \emph{Proceedings of the 2012 IEEE International Symposium on
  Information Theory}, pp. 566--570, July 2012.

\bibitem{HU13}
S.~H. Hassani and R.~Urbanke, ``Universal polar codes,'' July 2013,
  arXiv:1307.7223.

\bibitem{MELK132}
H.~Mahdavifar, M.~El-Khamy, J.~Lee, and I.~Kang, ``Compound polar codes,''
  February 2013, arXiv:1302.0265.

\bibitem{WG13}
M.~M. Wilde and S.~Guha, ``Polar codes for classical-quantum channels,''
  \emph{IEEE Transactions on Information Theory}, vol.~59, no.~2, pp.
  1175--1187, February 2013, arXiv:1109.2591.

\bibitem{WG12}
S.~Guha and M.~M. Wilde, ``Polar coding to achieve the {Holevo} capacity of a
  pure-loss optical channel,'' in \emph{Proceedings of the 2012 IEEE
  International Symposium on Information Theory}, Cambridge, MA, USA, July
  546-550.

\bibitem{WR12}
J.~M. Renes and M.~M. Wilde, ``Polar codes for private and quantum
  communication over arbitrary channels,'' \emph{IEEE Transactions on
  Information Theory}, vol.~60, no.~6, pp. 3090--3103, June 2014,
  arXiv:1212.2537.

\bibitem{WG13QDeg}
M.~M. Wilde and S.~Guha, ``Polar codes for degradable quantum channels,''
  \emph{IEEE Transactions on Information Theory}, vol.~59, no.~7, pp.
  4718--4729, July 2013, arXiv:1109.5346.

\bibitem{WLH13}
M.~M. Wilde, O.~Landon-Cardinal, and P.~Hayden, ``Towards efficient decoding of
  classical-quantum polar codes,'' \emph{Proceedings of the 8th Conference on
  the Theory of Quantum Computation, Communication and Cryptography (TQC
  2013)}, pp. 157--177, May 2013, arXiv:1302.0398.

\bibitem{RDR12}
J.~M. Renes, F.~Dupuis, and R.~Renner, ``Efficient polar coding of quantum
  information,'' \emph{Physical Review Letters}, vol. 109, no.~5, p. 050504,
  August 2012, arXiv:1109.3195.

\bibitem{BBN08}
\BIBentryALTinterwordspacing
I.~Bjelakovi\'c, H.~Boche, and J.~N\"otzel, ``Quantum capacity of a class of
  compound channels,'' \emph{Physical Review A}, vol.~78, no.~4, p. 042331,
  October 2008, arXiv:0808.1007. [Online]. Available:
  \url{http://link.aps.org/doi/10.1103/PhysRevA.78.042331}
\BIBentrySTDinterwordspacing

\bibitem{BB09}
I.~Bjelakovi\'c and H.~Boche, ``Classical capacities of averaged and compound
  quantum channels,'' \emph{IEEE Transactions on Information Theory}, vol.~55,
  no.~7, pp. 3360--3374, July 2009, arXiv:0710.3027.

\bibitem{STA09}
E.~\c{S}a\c{s}o\u{g}lu, E.~Telatar, and E.~Arikan, ``{Polarization for
  arbitrary discrete memoryless channels},'' August 2009, arXiv:0908.0302.

\end{thebibliography}

\begin{IEEEbiographynophoto}{Christoph Hirche}
received a BSc  degree in 2013 and a MSc degree in 2015 in Physics from the Leibniz Universit\"{a}t Hannover, Germany.
He is currently a PhD student at the Departament de F\'{i}sica, Universitat Aut\`{o}noma de Barcelona, Spain.
\end{IEEEbiographynophoto}

\enlargethispage{-6in}

\begin{IEEEbiographynophoto}{Ciara Morgan}
received an Honours BSc degree in Theoretical Physics and
Computer Science from the National University of Ireland, Maynooth in 2005.
In 2010 she was awarded a PhD in Mathematical Physics from University
College Dublin, Ireland, and was a postdoctoral fellow at the Centre for
Quantum Technologies, National University of Singapore until 2012. 
From 2012 to 2015 she was a postdoctoral researcher at the Institut f\"{u}r Theoretische Physik,
Leibniz Universit\"{a}t Hannover, Germany.
She is currently a Lecturer at the School of Mathematics and Statistics, University College Dublin, Ireland.
\end{IEEEbiographynophoto}

\begin{IEEEbiographynophoto}{Mark M. Wilde} (M'99 SM'13) was born in Metairie, Louisiana, USA.
He received the Ph.D. degree in electrical engineering from the University
of Southern California, Los Angeles, California, in 2008. He is an
Assistant Professor in the Department of Physics and Astronomy and the
Center for Computation and Technology at Louisiana State University. His
current research interests are in quantum Shannon theory, quantum optical
communication, quantum computational complexity theory, and quantum error
correction.
\end{IEEEbiographynophoto}

\end{document}